# Efficient Semiclassical Evaluation of Electronic Coherences in Polyatomic Molecules

Nikolay V. Golubev and Jiří Vaníček

*Abstract:* Exposing a molecule to intense light pulses may bring this molecule to a nonstationary quantum state, thus launching correlated dynamics of electronic and nuclear subsystems. Although much had been achieved in the understanding of fundamental physics behind the electron-nuclear interactions and dynamics, accurate numerical simulations of light-induced processes taking place in polyatomic molecules remain a formidable challenge. Here, we review a recently developed theoretical approach for evaluating electronic coherences in molecules, in which the ultrafast electronic dynamics is coupled to nuclear motion. The presented technique, which combines accurate *ab initio* on-the-fly simulations of electronic structure with efficient semiclassical procedure to compute the dynamics of nuclear wave packets, is not only computationally efficient, but also can help shed light on the underlying physical mechanisms of decoherence and revival of the electronic coherences driven by nuclear rearrangement.

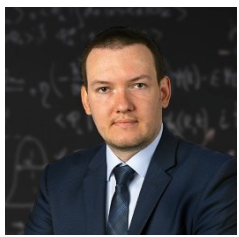

*Nikolay Golubev* studied Chemistry at Lomonosov Moscow State University, Russia, and obtained his PhD in Theoretical Chemistry at Heidelberg University, Germany. During his PhD studies, Nikolay was a fellow of International Max Planck Research School for Quantum Dynamics (IMPRS-QD) and received Dr. Sophie Bernthsen-Fonds Fellowship for extraordinary PhD students. In 2018, he was awarded a Branco Weiss Fellowship, and started as a junior group leader at Ecole Polytechnique Fédérale de Lausanne. His research focuses on developing and applying theory and simulation methods to investigate the ultrafast quantum dynamics in molecules and atomic clusters.

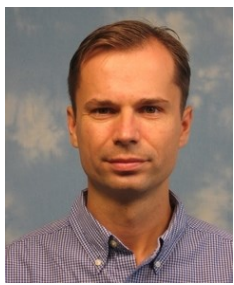

*Jiří Vaníček* earned his bachelor's degree in physics, summa cum laude, and PhD in theoretical physics from Harvard University. After postdoctoral positions at the University of California in Berkeley and at the Institute for Advanced Study in Princeton, Jiří Vaníček joined the faculty of the Department of Chemistry at Ecole Polytechnique Fédérale de Lausanne. He is the representative of Switzerland in the International Society for Theoretical Chemical Physics and the recipient of the OpenEye Outstanding Junior Faculty Award from the American Chemical Society and the Consolidator Grant from the European Research Council. His research focuses on developing efficient ab initio semiclassical and exact quantum dynamics methods with applications to steady-state and ultrafast spectroscopy.

———————

*Correspondence: Dr. N. Golubev, E-mail: nik.v.golubev@gmail.com, Prof. J. Vaníček, E-mail: jiri.vanicek@epfl.ch, Laboratory of Theoretical Physical Chemistry, Institut des Sciences et Ingénierie Chimiques, Ecole Polytechnique Fédérale de Lausanne (EPFL), Av. F.-A. Forel 2, CH-1015 Lausanne, Switzerland

## 1. Introduction

Photoinduced molecular processes play a key role in physics, chemistry, and biology. In nature, light triggers a large variety of chemical reactions such as photosynthesis, vision, and the formation of vitamin D, but also can cause the radiation damage of biomolecules and photolysis.[1] Furthermore, the interaction of light with matter forms the basis of important technological applications such as solar cells in which photoinduced charge transfer and light harvesting are essential.[2] What all these processes have in common is capturing the energy of light and its transformation into other forms of energy like heat, electricity, or chemical energy. On a microscopic level, this energy conversion process is the result of a correlated motion of electrons and nuclei after photoexcitation or photoionization of a molecule. Therefore, understanding of fundamental principles lying behind the light-induced electronic and nuclear dynamics in molecules is of crucial importance to comprehend the diversity of the world or even life.

Unraveling the ultrafast motion of electrons and nuclei in molecules is an enormously difficult task since the timescale of this movement is extremely rapid. As a case in point, the molecular vibrations typically occur on a timescale of tens to hundreds of femtoseconds (1 fs = $10^{-15}$ s) which corresponds to the atomic movement with the speed ≈ 1 km/s, while the motion of electrons takes place on even faster, attosecond (1 as = $10^{-18}$ s), timescale. Clearly, experimental observations of such an ultrafast molecular dynamics require the usage of advanced techniques with very high temporal and spatial resolution.

With the advent of the twenty-first century, developments of coherent light sources enabled the creation of subfemtosecond laser pulses with remarkably well controlled parameters.[3] Ultrashort intense laser pulses revolutionized the field of atomic and molecular physics providing the scientific community with a unique tool to initiate and trace dynamics of both electrons and nuclei of a molecule in real time and with atomic spatial resolution. This progress made it possible to study the fundamental concepts such as electron-nuclear correlation, charge transfer and migration, electronic coherence as well as its decoherence and revivals due to nuclear motion.

In particular, recent experiments utilizing a sequence of an isolated attosecond pulse and an intense few-cycle IR pulse have demonstrated the possibility to resolve the correlated electron-nuclear dynamics during the dissociative ionization of $H_2$ and $D_2$ molecules,[4] as well as to measure similar processes in other more complicated diatomic molecules.[5,6] In pioneering experiments of Calegari and co-workers, evidence of ultrafast charge migration in amino acids phenylalanine[7,8] and tryptophan[9] was demonstrated by measuring the yield of a doubly charged ion as a function of the delay between the ionizing pump and doubly ionizing probe pulses. In the seminal work by Kraus et al.,[10] attosecond charge migration in ionized iodoacetylene was reconstructed and controlled by analyzing the high-harmonic generation (HHG) spectrum emitted after irradiation of the molecule with strong infrared pulses of different wavelengths (800 and 1300 nm). Very recently, the decoherence and revival of ultrafast charge oscillations driven by non-adiabatic dynamics in neutral silane has been observed using X-ray attosecond transient absorption spectroscopy (ATAS).[11]

Although the development of advanced experimental techniques has produced a wealth of data, providing detailed information on the dynamics and properties of molecules, very often the interpretation of the experimental results cannot be done without sophisticated theoretical simulations. For this reason, the role of theory as a prerequisite to outline new experiments, and to disclose their feasibility and informational content, becomes stronger than ever before.[12]

In this brief overview, we discuss how the electronic coherence and decoherence can be analyzed and interpreted with a recently developed theoretical approach which combines accurate «on-the-fly» ab initio calculations of electronic structure with efficient semiclassical procedure to describe the dynamics of nuclear wave packets. We show that this simple single-trajectory scheme can evaluate the electronic coherence time in polyatomic molecules accurately by demonstrating an excellent agreement with full-dimensional quantum calculations. Next, we explain how the presented technique was, due to its computational efficiency, employed to scan a large variety of molecules searching for new systems with a long-lasting oscillatory dynamics of the electron density. Then we present an application of the method to unravel the physical mechanism of decoherence and of the revival of ultrafast charge oscillations in silane. Finally, we discuss briefly possible improvements of the presented scheme which can make on-the-fly semiclassical simulations more accurate.

## 2 Theoretical description of electron-nuclear dynamics in molecules

Exposing a molecule to intense light pulses, one creates a superposition of several electronic and vibrational quantum states. To describe the evolution of such initial superposition in time, one needs to solve the quantum mechanical equation of motion which is known as the time-dependent Schrödinger equation. Although nearly hundred years have passed since the discovery of this equation, such simulations remain extremely difficult for systems with more than a few degrees of freedom. In all other cases of interest, one needs to introduce approximations in order to simplify the mathematical treatment of quantum dynamics in systems containing many interacting particles.

A common starting point for most approaches to simulate the dynamics of molecular systems is the Born–Oppenheimer–Huang (BOH) expansion of the molecular wavefunction.[13,14] This expansion exploits the fact that atomic nuclei are much heavier than electrons, which makes it possible to separate the fast electron motion from the usually much slower motion of the nuclei and to represent molecular dynamics as the dynamics of nuclei moving on the potential energy surfaces (PESs) formed by the electrons in a specific electronic state. Importantly, the BOH approach is formally exact if the set of electronic states used in the expansion of the molecular wavefunction is complete. It is only when the expansion is truncated that an approximation is introduced.[15]

In the BOH picture, the electron dynamics occurs as a result of the coherent superposition of multiple stationary electronic states:

$$\Psi(r, R, t) = \sum_I \chi_I(R, t) \Phi_I(r, R), \quad (1)$$

where $\chi_I(R, t)$ is the time-dependent nuclear wave packet propagated on the PES associated to the $I$-th electronic state $\Phi_I(r, R)$, and $r$ and $R$ denote electronic and nuclear coordinates, respectively. The electronic states, together with the corresponding PESs $V_I(R)$, are obtained by solving the stationary Schrödinger equation for the electronic subsystem for all possible nuclear configurations of a molecule. A major challenge in this approach is to describe the quantum mechanical behavior of nuclear wave packets in many dimensions and in the presence of nonadiabatic effects caused by couplings betweeen different electronic states.

One of the most powerful approaches to an accurate description of the nuclear dynamics taking place on the manifold of electronic states is the multi-configurational time-dependent Hartree (MCTDH) method.[16] MCTDH has been used successfully for simulating nonadiabatic wave packet dynamics in molecules[17] and recently applied to the description of coherent electron-nuclear motion[18,19]. Although this rigorous technique makes it possible to take into account all the quantum effects, such as tunneling and nonadiabatic transitions, it suffers from an exponential scaling problem and also requires the costly construction of global PESs.

An alternative strategy for simulating coupled electron-nuclear dynamics employs a trajectory-guided Gaussian basis to represent the evolving wave packet and an «on-the-fly» evaluation of the electronic structure. These «direct dynamics» approaches calculate the PESs along trajectories only, thus avoiding the precomputation of globally fitted surfaces, and sample only the relevant regions of the configuration space. Among these methods, the closest in spirit to MCTDH are the variational multi-configurational Gaussians (vMCG)[20–23], but many others exist, ranging from *ab initio* multiple spawning[24] and other Gaussian basis methods[25–27] to more approximate mixed quantum-classical[28–30] and semiclassical[31–33] approaches.

### 2.1 *Time-dependent observable properties*

To analyze and visualize the electron motion in a system, an observable property must be computed. Utilizing the BOH form of the molecular wavefunction, Eq. (1), the expectation value of a general electronic operator $\hat{O}(r, R)$ can be written as

$$\begin{aligned}\langle \hat{O} \rangle(t) &= \iint \Psi^*(r, R, t) \hat{O}(r, R) \Psi(r, R, t) dr dR \\ &= \sum_{I,J} \int \chi_I^*(R, t) O_{IJ}(R) \chi_J(R, t) dR, \end{aligned} \quad (2)$$

where $O_{IJ}(R)$ are matrix elements of the $\hat{O}(r, R)$ operator between electronic states $I$ and $J$. If both the operator $\hat{O}(r, R)$ and the electronic states $\{\Phi_I(r, R)\}$ depend weakly on nuclear coordinates $R$, Eq. (2) can be further simplified as

$$\langle \hat{O} \rangle(t) \approx \sum_{I,J} O_{IJ} \chi_{IJ}(t), \quad (3)$$

where the coefficients

$$\chi_{IJ}(t) = \int \chi_I^*(R, t) \chi_J(R, t) dR \quad (4)$$

represent the populations of electronic states when $I = J$ and the electronic coherences when $I \neq J$. Being the only time-dependent quantities in Eq. (3) for the expectation value, the populations and coherences can serve as convenient properties to analyze the evolution of observables of a system in time.

### 2.2 *Direct dynamics methods for computing electronic coherence*

Treating the electronic dynamics coupled to nuclear motion with trajectory-based techniques was pioneered by Bearpark, Robb and co-workers[34–37] who propagated multiple trajectories representing an initially delocalized wave packet using the mean-field Ehrenfest approximation. Although this technique captures decoherence due to the destructive interference of coherent oscillations with different frequencies along the different trajectories, it completely ignores the decoherence due to the decay of overlaps of the nuclear wavepackets moving on different PESs. Avoiding the mean-field approximation by allowing the nuclear wave packets launched on different PESs to evolve independently, the decay of nuclear overlaps was taken into account and the electronic coherence upon ionization of a system was simulated in several molecules using the direct dynamics versions of vMCG scheme[38,39].

Recently, it was demonstrated that a simple semiclassical approach, in which the evolving nuclear wave packet is approximated by a single Gaussian function, can compete in accuracy with the full-dimensional quantum techniques for calculating the electronic coherence.[40] In this semiclassical approach, originally proposed by Heller,[41] the center of the Gaussian follows classical Hamilton's equations of motion while the width and the phase of the wave packet are propagated using the local harmonic approximation of the PES. Because the width of the Gaussian evolves in time, Heller's

method has been referred to as the "thawed Gaussian approximation" (TGA) in the literature.[42–44]

The Gaussian wave packet considered in the TGA is given in the position representation as

$$\chi_I(\mathbf{R}, t) = \sqrt{p_I} \exp\left\{\frac{i}{\hbar}\left[\frac{1}{2}\left(\mathbf{R} - \mathbf{R}_t^I\right)^T \cdot \mathbf{A}_t^I \cdot \left(\mathbf{R} - \mathbf{R}_t^I\right) + \left(\mathbf{P}_t^I\right)^T \cdot \left(\mathbf{R} - \mathbf{R}_t^I\right) + \gamma_t^I\right]\right\}$$

(5)

where $p_I$ is the population of the corresponding electronic state, $\mathbf{R}_t^I$ and $\mathbf{P}_t^I$ are the phase-space coordinates of the center of the wave packet, $\mathbf{A}_t^I$ is a complex symmetric width matrix with a positive-definite imaginary part, and $\gamma_t^I$ is a complex number whose real part is a dynamical phase and imaginary part ensures the normalization at all times.

The simple Gaussian form of the nuclear wave packet (5) imposed by the TGA makes it possible to perform the integration step in Eq. (4) for the electronic coherences analytically.[45] To simplify the resulting expression, we assume, in addition, that the widths of the Gaussians propagated in the electronic states $I$ and $J$ remain fixed along trajectories (this is often called the "frozen Gaussian approximation"), thus obtaining:[40]

$$\chi_{IJ}(t) \approx \sqrt{p_I p_J}\, e^{-\frac{\Delta R_{IJ}(t)^2}{4\hbar}} e^{-\frac{\Delta P_{IJ}(t)^2}{4\hbar}} e^{-\frac{iS(t)}{\hbar}}.$$

(6)

Here, $\Delta R_{IJ}(t) = |\mathbf{R}_t^I - \mathbf{R}_t^J|$ and $\Delta P_{IJ}(t) = |\mathbf{P}_t^I - \mathbf{P}_t^J|$ are the spatial and momentum separations, respectively, between the centers of the two Gaussian wave packets at time $t$ in mass- and frequency-scaled phase-space coordinates $\mathbf{R}$ and $\mathbf{P}$, and

$$S(t) = \Delta E t - \oint \mathbf{P}^T \cdot d\mathbf{R},$$

(7)

is the classical action, where the first term $\Delta E t$ is reponsible for electronic oscillations with the frequency corresponding to the energy difference between the involved electronic states $I$ and $J$ at the stationary geometry $\mathbf{R}_0$, and the second term is the reduced action quantifying the difference in frequency due to the divergent motion of wave packets on PESs $I$ and $J$.

The analytical expressions (6) and (7) permit a simple semiclassical interpretation of the effect of nuclear dynamics on electronic coherence. The decay of coherence takes place due to the increasing distance between the nuclear wave packets in phase space. Besides the influence on the absolute value of the electronic coherence, the diverging nuclear trajectories affect the frequency of electronic oscillations. In the absence of nuclear motion, the time scale of electronic dynamics is defined by the energy difference between the corresponding electronic states. Due to the nuclear motion, the static frequency is modified by the signed phase-space area between two nuclear trajectories (see Ref.[40] for more details).

## 3 On-the-fly *ab initio* simulations of electronic coherences in polyatomic molecules

Let us now consider the quantum dynamics triggered by ionization of a molecule with an ultrashort laser pulse. We assume that the duration of the applied pulse is shorter than the electron correlation, i.e. the ionization event takes place on an infinitely short time scale, which allows us to employ the so-called sudden approximation.[46,47] In this case, the initial ionic state is prepared by projecting the electronic wavefunction obtained by removing an electron from the ground neutral state of a molecule onto the ionic subspace of the system.

The created initial state is, in general, not an eigenstate of the cationic Hamiltonian and, thus, will evolve in time. The ultrafast multielectron dynamics triggered by the ionization can manifest as the time-evolution of the created hole along a molecular chain. Driven by purely electronic effects, this mechanism was termed "charge migration"[48,49] to distinguish it from a more common "charge transfer" driven by nuclei[50,51].

A showcase example where the ultrafast ionization initiates the charge migration oscillations is the propiolic acid ($HC_2COOH$). This molecule provides a perfect system for validating the semiclassical TGA because the electronic coherences in this system were recently calculated using a full quantum MCTDH approach.[19]

While dynamically exact, the MCTDH simulations employed an approximate, vibronic-coupling (VC) Hamiltonian model[15] to represent the PESs of the involved electronic states. *Ab initio* third-order algebraic diagrammatic construction [ADC(3)] method[52] with the standard double-zeta plus polarization (DZP)[53] basis set was used to compute the corresponding PESs. Assuming the ionization out of the highest occupied molecular orbital (HOMO), a coherent superposition of the first and the third cationic states of the molecule was created at the itial moment of time according to the corresponding hole-mixing weights[54] and employing the sudden ionization mechanism.

The full quantum-mechanical calculations show (red solid line in Fig. 1) that the electronic oscillations are strongly influenced by nuclear motion—the coherences are completely suppressed within first 15 fs.[19] To estimate the importance of nonadiabatic dynamics, we repeated the quantum calculation on a reduced, "diabatic" version of the VC model, where the diabatic PESs are not coupled to each other. In this case, the TGA is exact and therefore provides results identical to the MCTDH simulations (yellow solid line in Fig. 1). The small deviations between the full and reduced versions of VC model start to appear at longer times due to the nonadiabatic effects.

We also performed semiclassical simulations using adiabatic PESs obtained by diagonalizing the diabatic VC model (blue dashed line in Fig. 1). Although in this case the adiabatic form of the Hamiltonian does not provide additional insights about the dynamics of nuclear wave packets, it plays a role of an intermediate step between computations performed using the precomputed PESs and those performed with the exact PESs obtained on the fly along the trajectories. Allowing the wave packet to evolve according to the exact Hamiltonian computed on the fly makes it possible to visit nuclear regions inaccessible within the VC Hamiltonian and thus to take anharmonicity effects into account. This is reflected in our on-the-fly calculations (green dash-dotted line in Fig. 1), which predict the electronic motion with a similar oscillation period, but a slightly faster decay of the electronic coherence than within the VC model. Remarkably, because the effect of using the on-the-fly potential is much larger than the effect of including the nonadiabatic couplings, the semiclassical on-the-fly result of the TGA is most likely more accurate than the quantum result of the MCTDH calculation with the VC model!

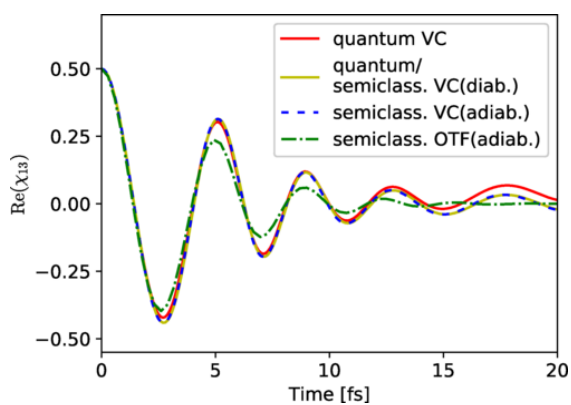

*Fig. 1. Comparison of various techniques for computing the electronic coherence measured by the time-dependent overlap $\chi_{13}(t)$ of the nuclear wave packets propagated on the first and third cationic states of propiolic acid after removal of the HOMO electron.*

Simulations of electronic coherences with the semiclassical TGA require the propagation of a single nuclear Gaussian wave packet on each involved PES, while the wave packets moving in different states are independent from each other. The computational cost of the employed scheme is comparable to the conventional *ab initio* molecular dynamics (in particular, the costs are the same if the width of the Gaussian is fixed). Therefore, in this setting, the TGA not only demonstrates an accuracy comparable to full-dimensional quantum methods, but can also be very computationally efficient.

Encouraged by this observation, we applied the TGA technique to perform a massive scan of small polyatomic molecules searching for specific structural and dynamical properties which can be useful for the experimental measurements of ultrafast electronic dynamics and its coupling to the nuclear motion. We used a free online database PubChem maintained by the National Institute of Health to analyse the correlated electron-nuclear dynamics in about 250 randomly selected small molecules composed of C, H, O, and N atoms.

Our simulations demonstrate (see Ref.[55] for more details) that the electronic coherences created after the ultrafast ionization become damped by the nuclear rearrangement on a time scale of a few femtoseconds in most of the studied molecules. However, we were able to identify several so far unexplored molecules with electronic coherences lasting up to 10 fs, which are, therefore, promising candidates for future experimental studies.[55]

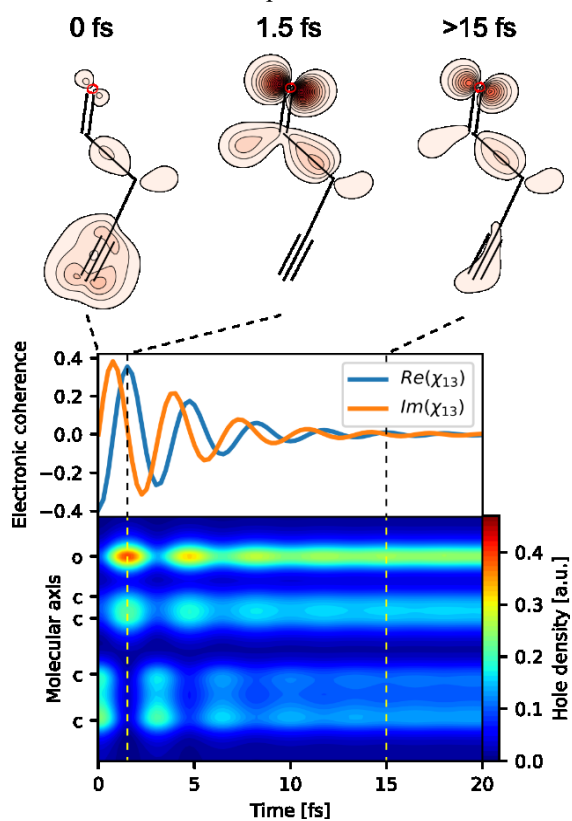

*Fig. 2. Correlated electron-nuclear dynamics triggered by the ionization out of the HOMO of the but-3-ynal molecule. Top: Snapshots showing the dynamics of the created positive charge along a molecular chain. Middle: Time evolution of the electronic coherence between the first and the third cationic states. Bottom: Time evolution of the hole density along the molecular axis. The charge initially localized in the HOMO orbital migrates back and forth between alkyne and aldehyde moieties of the molecule before being trapped by the nuclear motion.*

Figure 2 shows electronic coherences and the dynamics of a positive charge created after sudden ionization of the HOMO electron of the but-3-ynal molecule. As one can see, the hole, initially localized around the carbon triple bond of the molecule, starts to oscillate between the triple bond and the aldehyde group with a period of about 3.8 fs, determined by the energy gap between the involved cationic states. The charge performs several clear oscillations before being trapped by the nuclear motion and distributed along the molecular chain.

We would like to point out also the obvious limitations of the presented methodology. It is well known that the TGA cannot capture tunnelling and wave-packet splitting, but these effects are typically unimportant at the ultrashort time scale on which decoherence occurs. In the present context, the main weakness of the TGA is its inability to take nonadiabatic transitions into account. In our simulations, we concentrated only on those molecules in which the energy gaps between the ionic states remain large enough to justify the negligibility of the nonadiabatic population transfer. The presence of conical intersections in the vicinity of the created wave packets can dramatically alter the subsequent dynamics, and thus, the usage of more sophisticated approaches for molecular dynamics is required in this case. Nevertheless, the ultrafast electron dynamics studied in this work occurs on sub-femtosecond time scale which typically implies that the nuclei remain very close to their original positions and often do not have enough time to reach regions of the configurational space where the nonadiabatic couplings are strong. Unless the excitation or ionization places the wave packet directly at a conical intersection, the TGA agrees with the exact solution of the time-dependent Schrödinger equation in the limit when the propagation time approaches zero and, thus, is particularly suited for the treatment of processes taking place on ultrashort time scales.

## 4 Revealing the physical mechanism of decoherence

As we discussed already in Sec. 2.2, the semiclassical Gaussian method makes it possible to reveal the physical mechanism of decoherence beyond the simple but vague blame on nuclear motion. Here, we apply an extension of the technique developed in Ref.[40] to interpret the recent experimental measurements of the decoherence and revival of the ultrafast charge oscillations in neutral silane.[11] The experiment, performed by the group of H. J. Wörner at ETH Zürich, utilizes an intense infrared 5.2-fs laser pulse to excite the molecule and an isolated sub-200-as X-ray pulse to probe the induced dynamics using the ATAS technique.[11] It has been shown[56] that the major ingredient required to simulate the ATAS is the electronic coherences. We therefore concentrate here on analyzing the electronic coherences between the excited states of a neutral silane created after the interaction of the molecule with the pump pulse.

*Ab initio* computations of the valence excited electronic states of silane demonstrated[11] that the corresponding PESs have both non-adiabatic and Jahn–Teller interactions which cannot be taken into account within the discussed semiclassical scheme. Therefore, fully quantum MCTDH method with a second-order VC Hamiltonian was used to reproduce the experimentally measured electronic coherences (see the Supplemental Information of Ref.[11] for details). The simulations revealed that the experimentally measured signal results from the superposition of a pair of electronically excited states, which we shall denote simply as $I$ and $J$. Even though the results of MCTDH calculations agree very well with the experimental measurements[11] and it is clear that the nuclear dynamics is responsible for the modulation of the coherence, the internal mechanism of the process is hard to deduce from sophisticated numerical quantum methods such as MCTDH.

To understand the mechanism of dephasing and revival of the electronic coherence observed in the experiment and reproduced by the accurate MCTDH simulations, we performed a theoretical analysis using a simple Eq. (6) for the overlap of Gaussian wave packets. Because our scheme cannot take the nonadiabatic transitions into account, we do not run the semiclassical propagation explicitly as we did in Ref.[40]. Instead, we analyze the fully quantum MCTDH wave packets dynamics using the formalism presented in Sec. 2.2.

To establish the correspondence between the quantum wave packet and its semiclassical representation, we set the position $\boldsymbol{R}_t^I$ and momentum $\boldsymbol{P}_t^I$ of the Gaussian by computing the expectation values of the corresponding operators for the quantum wave packet evolving in the electronic state $I$:

$$\boldsymbol{R}_t^I = \int \chi_I^*(\boldsymbol{R},t) \boldsymbol{R} \chi_I(\boldsymbol{R},t) d\boldsymbol{R},$$

$$\boldsymbol{P}_t^I = -i\hbar \int \chi_I^*(\boldsymbol{R},t) \nabla \chi_I(\boldsymbol{R},t) d\boldsymbol{R}.$$

(8)

Extracting in the same way the position and momentum of the nuclear wave packet evolving in the electronic state $J$, we can use Eqs. (6) and (7) to analyze the electronic coherences. We will focus on the magnitude of the electronic coherence, i.e., on the quantity $|\chi_{IJ}(t)|$, and will not discuss the phase $S(t)/\hbar$.

Figure 3 illustrates the MCTDH-calculated electronic coherence (panel A in Fig. 3) and the semiclassical analysis which elucidates the origin of the decay and revival of the electronic coherence between states $I$ and $J$. In order for coherence between a pair of the electronic states to exist, the nuclear wave packets evolving in these states must overlap in both coordinate and momentum space (see panel B in Fig. 3 for the spatial and momentum separations between the wave packets, and panel C in Fig. 3 for the corresponding contributions to the coherence). The comparison of the normalized magnitude of the electronic coherence (panel D in Fig. 3) demonstrates a good agreement between fully quantum MCTDH simulations and the semiclassical results. The semiclassical coherence slightly overestimates the quantum one because the evolving wave packets do not conserve their Gaussian form during their propagation on the corresponding PESs.

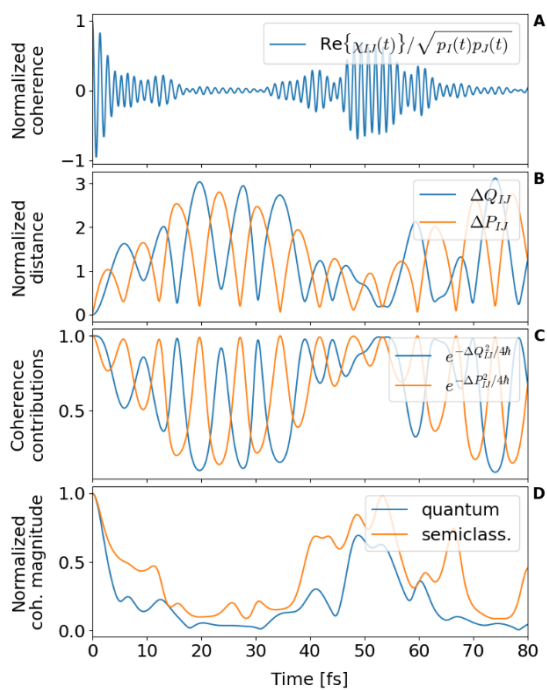

*Fig. 3. Semiclassical analysis of the electronic coherence created after the excitation of the neutral silane by a laser pulse. (A): Normalized electronic coherence computed by fully quantum MCTDH method; (B): difference between the position (or momentum) expectation values of the wave packets in the I and J states; (C): spatial and momentum contributions to the decay of semiclassical coherence [Eq. ( 6 )]; (D): comparison of the normalized magnitudes of the electronic coherence computed by quantum MCTDH and semiclassical Gaussian techniques.*

**5 Conclusions and Outlook**

In conclusion, we presented a simple and efficient on-the-fly semiclassical approach for computing electronic coherences in polyatomic molecules. We also described applications of this method to analyze the effects of nuclear motion on the electronic coherence in the propiolic acid and the decoherence and revival of the coherence in silane. On one hand, the remarkable computational efficiency of the method made it possible to scan a large number of polyatomic molecules searching for systems with the long-lasting oscillatory electronic dynamics. On the other hand, the simplicity of the semiclassical method (in comparison with sophisticated wave packet grid-based quantum methods) permitted us to disentangle several physical mechanisms contributing to decoherence and to the revival of coherence.

The semiclassical TGA used in this paper has a plenty of room for improvement. Extensions of the TGA, such as the extended thawed Gaussian approximation[57,58] or Hagedorn wave packets[43,59], which propagate a Gaussian multiplied by a linear or general polynomial, can make on-the-fly semiclassical simulations even more accurate. Furthermore, extensions of the TGA scheme to include the finite temperature effects into consideration have been reported in recent years.[45]

Finally, we would like to point out that the efficient approach used in this work opens the door to the analysis of electron-nuclear dynamical processes in larger, biologically relevant systems. Being able to treat molecules with a few hundred atoms, the TGA technique combined with the appropriate electronic structure method can help shed light on the continuing debates on the role of quantum coherence in biology,[60–62] quickly preselect molecules suitable for further experimental investigations, and support theoretically recent experimental observations of attosecond electron dynamics in realistic molecular systems. We hope that our work will motivate such studies.

*Acknowledgements*
The authors are thankful to Victor Despré and Alexander Kuleff for sharing parameters of the vibronic-coupling Hamiltonian of silane and acknowledge financial support from the Swiss National Science Foundation through the National Center of Competence in Research MUST (Molecular Ultrafast Science and Technology) (phase III) (51NF40-183615) and from the European Research Council (ERC) under the European Union's Horizon 2020 Research and Innovation Programme (Grant Agreement No. 683069–MOLEQULE). N.V.G. acknowledges support by the Branco Weiss Fellowship—Society in Science, administered by the ETH Zürich.

Received: xx.04.2022